# Molecular stochastic process on gold surface observed in broadband-infrared, background-suppressed, sum frequency generation spectroscopy: picosecond heat transfer to self-assembled monolayers

*Hiroki Fujiwara*[*†]

Department of Chemistry, University of Illinois at Urbana-Champaign, 600 South Mathews Avenue, Urbana, Illinois 61801, USA

ABSTRACT. The picosecond thermal response of a normal octadecanethiol self-assembled monolayer on a gold surface was studied using pump-probe, broadband-infrared, background-suppressed and vibrational sum frequency generation spectroscopy. The orientation fluctuations in response to flash heating were characterized from the intensity kinetics of the terminal methyl C-H stretching vibrations. An intensity decrease (5.7 +/- 1.3 ps) in the asymmetric modes indicated that stochastic processes of molecular conformations

---

[*]Present addresses: Department of Chemistry, Durham University, South Road, Durham DH1 3LH, United Kingdom

[†]E-mail address: hiroki.fujiwara@durham.ac.uk, Phone number: +44 (0)191 334 2596

developed during multiple light-matter interactions. Including torsional diffusion into the rotation term of second order electrosusceptibility explained the experimental surface temperature increment two order-of-magnitudes smaller than the activation energy of the dihedral angles (~15 K).

**Introduction**

This paper describes the use of femtosecond-laser irradiation on a metal surface to examine the real-time response of self-assembled monolayers (SAMs) on a metal surface.[1] The pulse energy is received mainly by electrons due to their smaller heat capacity relative to phonons (two temperature model) [2,3]. The hot electrons in a thin metal film reaches 1000s K within the pulse duration and dissipates the excess energy to phonons in few ps, if not the hot electrons diffuse outside of the irradiated area by themselves.[2,3] The subsequent heat diffusion to a supporting substrate takes longer than few ns.[4] In characterizing the internal state of the adsorbate, we used multiplex, vibration-resonant sum frequency generation (SFG) spectroscopy (Fig 1a).[5] Previously, the same methods were used to investigate the disorder of a monolayer structure by observing the terminal C-H stretching fundamentals of normal-alkylthiol SAMs.[4,6-8] However, this disordering dynamics has not been analyzed in detail, in spite of the fact that SFG spectroscopy is widely used for obtaining molecular orientations or order parameters of interfaces.[9-11] One reason for this may be stochastic processes of molecular conformations that develop between the two laser pulses of broadband infrared SFG. It is an assumption of molecular orientation motions being significantly slower than the optical processes (slow limit) that a static ensemble average of molecular dipole and polarizability orientations in relation to the probe electric fields or direct cosines represents vibrational SFG polarization.[12,13] The vibrational SFG polarization is described by a dipole-polarizability response function in time domain.[14] However, to the best of our knowledge, only a few studies have discussed the time-correlation effects of the molecular orientations in the response function.[15] In this study, we performed a quantitative study on the intensity kinetics of the doubly degenerate C-H asymmetric modes ($v_a$) of the terminal methyl group in an n-octadecanethiol (ODT) SAM[16] and explained the behavior as its torsional motions, apart from the well-known kinetics of the symmetric mode ($v_s$).[17]

**Materials and Methods**

The apparatus used for time-resolved SFG spectroscopy was similar to that previously reported (Fig. 1(a)).[4,6,18] A femtosecond Ti:sapphire pulse-laser (~1.5 W, 1 kHz) was used as the light source. The probe was composed of broadband mid-infrared ($\omega_{IR}$, 3.4 μm, 200 fs, FWHM 300 cm$^{-1}$) and narrowband "visible" ($\omega_{vis}$, 800 nm, ~1.5 ps, FWHM 7 cm$^{-1}$) pulses, the beam paths of which were focused and crossed on the SAM at a 60° angle in pp polarization combinations. We used a background suppression technique which eliminates the non-resonant component of the SFG signal originating from the gold (Au) surface (Fig. 1(b)).[14,18,19] The $\omega_{vis}$ pulse was delayed "$\tau$" from $\omega_{IR}$ pulse, until only the resonant transition of $\omega_{IR}$ remained the coherence between ground and excited states.[15] A spectrometer and charge-coupled device were used to detect the SFG signal without the selection of polarizations. The pump pulse ($\omega_{pump}$, 800 nm, ~160 fs) was incident from the underside of the metal layers to avoid direct excitation of the monolayer (Fig. 1 (a)). The $\omega_{vis}$ pulse was delayed by 0.7 ps ($\tau$) after the $\omega_{IR}$ pulse. Spectra without the pump pulse, $\omega_{pump}$, were also recorded at every delay time ($t$) and scan as a reference or negative control. The $\omega_{pump}$ pulse energy was maximized such that the reference SFG intensity and spectral shape were unchanged during the experiment. The phonon temperature increase $\Delta T$ was ~15 K, estimated from the transient Drude reflectance from the Au surface and the $\nu_s$ SFG intensity change,[20] as described elsewhere.[14] We newly measured the reflection spectra from a temperature-controlled clean Au surface and prepared a calibration curve between the reflectance change and $\Delta T$, while the former temperature evaluation was based on the assumption that the visual damage threshold is Au melting temperature (1337 K).[4,6,14,18] This is because excitation at 400 nm pulse showed a higher reflectivity at the damage threshold intensity. The beam size (1/e intensity ellipse) of the $\omega_{pump}$ pulse on the Au surface was 0.1 mm$^2$ with ca. 15 mJ/cm$^2$ of peak fluence. The probe beam sizes were ca. 1/4 of $\omega_{pump}$ pulse. The reproducibility was confirmed by repeating the experiment on another day. The SAMs were prepared by immersing a substrate overnight in 1 mM ODT solution of special-grade ethanol purged with argon. ODT (Aldrich, 98%) was used without further purification. The 50 nm Au (111) and adhesive (Cr, 0.8 nm) layers were deposited on a glass plate (25×25×1.6 mm$^3$) via electron beam evaporation and used as the substrate.

**Results**

A steady SFG spectrum is shown in Fig. 2(a). All three peaks, evident at first glance, originated from the terminal methyl group due to the quasi-inversion symmetry of methylene pairs;[4,6,18] (from short to long wavelength) C-H asymmetric stretching $\nu_a$, Fermi's resonance $\nu_F$, and C-H symmetric stretching $\nu_s$. We detected the absolute square of the second order electrosusceptibility ($\chi_2$) multiplied by the infrared probe pulse ($\omega_{IR}$) spectral shape $g(\omega_{IR})$:[17]

$$I(\omega_{SFG} = \omega_{IR} + \omega_{vis}) \propto g(\omega_{IR}) \left| A_{NR} e^{i\psi_{NR}} + \sum_n \frac{A_n e^{i\psi_n(\omega_{IR})}}{\omega_{IR} - \omega_n - i\Gamma_n} \right|^2, \qquad (1)$$

where the first term inside the parentheses is the non-resonant component of $\chi_2$ from the Au surface with real-number amplitude ($A_{NR}$) and phase ($\phi_{NR}$). The second term is the sum of all resonant components of $\chi_2$. These line shapes are Lorenz functions of amplitude ($A_n$) and width ($\Gamma_n$),[21] because the $\omega_{vis}$ time profile was shaped in an exponentially decreasing function for maximal overlap of the longer lived vibrational response of the molecule.[19] It is known that the background suppression technique modifies each phase $\psi_n$ in the SFG spectral function depending on its frequency $\omega_{IR}$.[14,22] However, these phase modulations were straightforward, when the non-resonant component was negligible:[11]

$$\psi_n - \psi_m = -(\omega_n - \omega_m)\tau + 2k\pi, \tag{2}$$

where $k$ is an arbitrary integer. Fresnel factors did not alter the phases[11] when we assumed that only one element (zzz) of $\chi_2$ has a significant contribution to the SFG signal. We applied this constraint to eq. 1 and optimized the parameters via the least square regression method (Fig. 2(a-b)). Two solutions fit the spectra equally well (Fig. 2(c-d)). Comparing the retrieved relative sign information on resonant amplitudes ($A_n$) for the first and second solutions (Fig. 2(e) and (f), respectively) with that (Fig. 2(g)) for the standard SFG spectrum (Fig. 2(h)) with $\psi_n = 0$ and $\psi_{NR} = \pi/2$, we concluded that the second solution was correct. We also found that another resonance peak on the longer wavelength side from the $\nu_s$ peak was necessary to reproduce the spectral shapes, which corresponded to the methylene C-H symmetric stretching mode ($\nu_{CH2}$).[4]

In kinetic analyses, the phases ($\psi_n$, $n$ = 1, 2, 3), $\omega_{IR}$ spectral shape (g($\omega_{IR}$)), and $\nu_{CH2}$ parameters ($A_4$, $\Gamma_4$, and $\psi_4$) were constrained to be identical at all time-steps (global fit; see Fig. 3a). We interpreted the area ($\propto |A_n|^2/\Gamma_n$) of the corresponding absolute-square Lorentzian term in eq. 1 as the intensities, assuming that $\omega_{vis}$ pulse duration was significantly shorter than the vibrational dephasing rate. The $\nu_s$ intensity decreased by 21±1% (standard deviation) from the initial intensities ($t \leqq -2$ ps), assuming double exponential kinetics starting from $t = 0$: $I_0 + A_1 \exp(-t/\tau_1) + A_2 \exp(-t/\tau_2)$. The faster time constant agreed with the previous report investigating a larger $\Delta T$, considering that the standard vibrational SFG signal is a heterodyne detection with non-resonant background from gold as a local oscillator:[4] 5.2 ± 0.7 (standard deviation) and 79±48 ps (Fig. 3(b)). The decrease (13±1%) in the $\nu_a$ intensity also agreed with the previously time-resolved spectra (Fig. 3(c)).[4] The time constant (5.7±1.2 ps) was identical to the faster $\nu_s$ component within the experimental error, assuming a single exponential function: $I_0 + A_1 \exp(-t/\tau_1)$. The intensities at $t = 0$ evaluated from the exponential fitting were already smaller than the initial values ($t \leqq -2$ ps) by 2.5±0.5 and 5±1 % for $\nu_s$ and $\nu_a$ modes, respectively, because of background suppression technique.[14]

**Discussion**

The kinetics in 5 to 6 ps time scale correspond to the monolayer orientation fluctuation, according to a physical picture of the ultrafast response of ODT SAM to a femtosecond laser presented by Dlott et al.[4] This fluctuating motions were successfully analyzed as a time-evolved probability density of the molecular orientation, in the former cases of shockwave excitations,[23] where the monolayer was immersed in a shock impedance matched liquid (deuterated ethylene glycol).[9,24] The orientation distribution of molecular hyperpolarizations ($\beta$) controls the ensemble average of resonant $\chi_2$ amplitudes ($A_n$) in the slow limit.[4] The methyl $\nu_a$ ($A_s$) and $\nu_s$ ($A_a$) modes on an interface of $C_{\infty v}$ symmetry depends on different order parameters:[9,10]

$$A_{zzz,s} \propto N_s \beta_{z'z'z'}[r\langle\cos\theta\rangle + (1-r)\langle\cos^3\theta\rangle] \tag{3}$$

$$A_{zzz,a} \propto N_s \beta_{z'x'x'}(\langle\cos\theta\rangle - \langle\cos^3\theta\rangle),$$

where $\theta$ and $r$ is the zenith angle between methyl $C_{3v}$ symmetry axis and normal to the metal surface (Fig. 4a inset) and depolarization ratio ($\beta_{x'x'z'}/\beta_{z'z'z'}$, ($x'$, $y'$, $z'$) are molecular coordinates) for the $\nu_s$ mode, respectively. Although the *xxz*, *xzx*, and *zxx* elements are also expected to contribute to the SFG signal in the case of our polarization combinations (ppp), the *x* element of the visible probe intensity is 5 and 0.3 percent of the *z* element at our incident angles, since the complex index of refraction for gold is ca. 0.18+$i$5.1 and 2.0+$i$21 at 800 nm and 3.4 μm in room temperature, respectively.[25,26] This ratio is not modulated significantly by a temperature increase of $\Delta T$ from room temperature (e.g. -0.006 percent at 800 nm).[27] When the average $\theta$ value is not far away (0 to 55°) from room-temperature equilibrium,[4] these equations indicate that $v_a$ intensity should increases when $v_s$ intensity decreases and vice versa, if the distribution of $\theta$ is negligible. A simultaneous decrease of the $v_s$ and $v_a$ intensities was previously explained by trans-to-gauche isomerization around C-C bonds (dihedral angles), which increases the deviation of $\theta$ values.[9,24] Our flash heating excitation is different from the shockwave experiments in that the external perturbation does not involve anisotropic volume change in picosecond time scale. According to a non-equilibrium molecular dynamics simulation, the standard deviation increase of the zenith angle $\theta$ standard deviation was 4° (16° to 20°) for a hexadecanthiol SAM, while the average $\theta$ value also increased 6° (38° to 44°) monotonically for the first 100 ps after the flash heating of a gold thin layer from 300 to 1073 K (ca. 80 percent of gold melting temperature).[7,8] These changes of the distribution cause -27 and +3.0 percent modifications in the SFG intensities of $v_s$ and $v_a$ modes, respectively. It is possible that we observed a larger deviation change in our experiment. As the author pointed out, the potential between gold and sulfate atoms might have a room for improvement. Actually, ODT SAM takes a pairing phase after annealing to 375 K, although we did not observe evolution of spectral shape or intensities during the experiments. However, the temperature increment $\Delta T$ in our experiment is ca. 2 percent of the simulation conditions. Assuming quasi-thermal equilibrium,[28,29] the increase of gauche-defect population in our experiment conditions should be ca. 5 percent of the simulation along the terminal dihedral angle, the coordinate of smallest-energy defect, according to the calculated potential energy surface of an isomerized ODT molecule among the all-trans monolayer (ca. 14 kJ/mol).[24] One of explanations that hot electron excitation accounts for the dominant part of heat transfer from gold surface to the monolayer, which is contrary to the conclusions of some literatures.[4,18] Certain molecular degrees of freedom could have higher excess energy than that corresponding phonon temperature rise in gold surface until intermolecular energy relaxation completes, if the intermolecular thermal conductivity and direct coupling of the monolayer to gold hot electrons has a significant magnitude compared with the gold electron-phonon coupling.

Another explanation is the rotational motion around terminal C-C bond (torsional angle $\lambda$, see Fig. 4(a), inset), which has no intermolecular steric hindrance and significantly smaller activation energy, as well as moment of inertia, than those of the dihedral motions.[30] A thermal fluctuation of $\lambda$ disorientates the $\omega_{IR}$-induced polarizations exclusively of $\nu_a$ mode and hinders its coherent interaction with $\omega_{vis}$ pulse, since $\nu_s$ and $\nu_a$ transition dipoles are parallel and perpendicular to methyl symmetry axis, respectively. When $\lambda$ motion controls the total dephasing rate (fast limit),[31] the random distribution of $\lambda$[9,10,24] eliminates all terms for the $\nu_a$ amplitude:[32]

$$A_{zzz,s} \propto N_s \beta_{z'z'z'}(\langle\cos^2\theta\rangle\langle\cos\theta\rangle + r\langle\sin^2\theta\rangle\langle\cos\theta\rangle) \tag{4}$$

$$A_{zzz,as} \propto N_s \beta_{z'x'x'}\langle\sin\theta\cos\theta\cos\lambda\rangle\langle\sin\theta\cos\lambda\rangle = 0.$$

Therefore, we consider the intermediate region,[15] in which the $\lambda$-induced dephasing rate competes with the vibrational dephasing rate, assuming the limit of impulsive probe pulses:

$$A_{zzz,s} \propto N_s \beta_{z'z'z'} \iint G(\Omega_f,\tau|\Omega_i)(\cos^2\theta_i\cos\theta_f + r\sin^2\theta_i\cos\theta_f)p_0(\Omega_i)d\Omega_f d\Omega_i \tag{5}$$

$$A_{zzz,as} \propto N_s \beta_{z'x'x'} \iint G(\Omega_f,\tau|\Omega_i)(\sin\theta_i\cos\theta_i\cos\lambda_i)(\sin\theta_f\cos\lambda_f)p_0(\Omega_i)d\Omega_f d\Omega_i,$$

where $\Omega_i$ and $\Omega_f$ are the orientations or $\{\theta,\phi,\lambda\}$ in Euler angles, where $\phi$ is the methyl azimuthal angle (see Fig. 4 inset), of the terminal methyl group when infrared absorption and stimulated Raman processes occur.[13] The probability density, $p_0(\Omega_i)d\Omega_i$ determines the orientation distribution at the thermal equilibrium, where we assumed the uniform distribution of $\phi$ and $\lambda$, i.e. $p_0(\Omega_i)d\Omega_i \approx p_0(\theta_i) \cdot (1/2\pi)^2 d\theta_i d\phi_i d\lambda_i$. The way these equations take the rotational diffusion into account is the conditional probability density $G(\Omega_f,\tau|\Omega_i)\,d\Omega_f d\Omega_i$, when $\omega_{vis}$ pulse arrives after the delay time $\tau$ from $\omega_{IR}$ absorption. Now we assume the slow limit of $\theta$, i.e. $G(\Omega_f,\tau|\Omega_i)d\Omega_f d\Omega_i \approx \delta(\theta_f -$ $\theta_i)G(\phi_f,\tau|\phi_i)G(\lambda_f,\tau|\lambda_i)d\theta_f d\phi_f d\lambda_f d\theta_i d\phi_i d\lambda_i$. The torsional dynamics of methyl groups or $G(\lambda_f,\tau|\lambda_i)d\lambda_f d\lambda_i$ was well investigated via nuclear magnetic resonance (NMR) at low temperature,[33] while the shape of $G(\phi_f,\tau|\phi_i)d\phi_f d\phi_i$ has no effects on our SFG signal. We evaluated the former by assuming that the torsional conformation of the methyl group takes only three minima on the potential energy surface, $\lambda = \lambda_i, \lambda_i \pm 2\pi/3$. The torsional motion is modeled as a Markov processes among the three minima, of identical rate coefficients whose temperature dependence were evaluated by Arrhenius formula (Site Jump model).[34] Arrhenius factor and activation energy for methyl groups are quoted from literature (0.17 ps inverse and 7.1 kJ/mol for thymidine, respectively).[35] A nearly 7% decrease in the $v_a$ intensity in eq. (5) (i.e. $\propto |A_{zzz,as}|^2$) amounted to an approximate 10 K increase in temperature in the Arrhenius formula (torsional temperature, Fig. 4(b)). This evaluation is valid even if $\omega_{vis}$ is not impulsive, as long as the $\omega_{vis}$ pulse duration being negligible relative to the total dephasing time constants. This value agrees with the phonon temperature increment $\Delta T$, which indicates that surface phonons, instead of electrons, dominated the interface heat transfer.[4,6,18]

    The slower component of $v_s$ intensity decrease was not observed previously, to best of our knowledge. This might indicate a higher sensitivity of the background-suppress method combined with our global fitting analysis. Since the results of the molecular dynamics simulation also indicated kinetics of both order parameters and temperatures well approximated by a single exponential function, this slower kinetics might represent a collective motion of monolayer molecules larger than the size of the simulation ensemble (75 molecules). The conformation change of a single monolayer molecule alters the potential energy surface for the conformations of the vicinity molecules, which shall cause the secondary thermalization process. On the other hand, we did not apply out curve fitting analyses to the early-stage

kinetics. During $-2 < t < 0$ ps, for example, the $\nu_a$ intensity (5±1 of 13±1 %) decreased more than the $\nu_s$ mode (2.5±0.5 of 21±1 %) in the ratios to the initial intensities. This difference is not explained by the change in reflectivity or Fresnel coefficients on gold surface due to two temperature model,[2,20] considering negligible refractive-index dispersion over the frequency range of ca. 90-cm$^{-1}$ (2.7-THz). One of the explanations of this dynamics might be hot-electron heating of the monolayer,[36] as we discussed previously regarding to benzene-linker SAMs.[18] Electron transfers could modify the torsional potential barrier, enhancing $\psi$ diffusions without excess energy in a sub-picosecond time scale.[3]

In conclusion, the time correlation of the torsion angle $\lambda$ rather than the probability distribution along the zenith angle $\theta$ characterized the intensity kinetics of methyl C-H asymmetric stretching mode $\nu_a$. We introduced a phase constraint (eq. 2) for the background-suppressed SFG spectral function and described direction cosines for $\chi_2$ in the case of intermediate methyl-top motion (eq. 4). The limitations of our method include the fact that hot-electron heating is completed (~1 ps) beyond our time resolution.[3] Since the visible narrow-band probe pulse $\omega_{vis}$ limited our time resolution, a good complementary approach would be to scale up the band widths in electron-transition frequencies. Even without a background suppression technique, heterodyne detection could separate the monolayer signals as the imaginary part of $\chi_2$ from the non-resonant SFG.[22,37]

**Acknowledgements**

This material is based upon Work-Made-for-Hire supported by US Air Force Office of Scientific Research under award number FA9550-09-1-0163, the Office of Naval Research under award N00014-11-1-0418, and the National Science Foundation under award DMR-09-55259 and prepared under the agreement of publication with the employer.

REFERENCES

1. Love JC, Estroff LA, Kriebel JK, Nuzzo RG, Whitesides GM (2005) Self-assembled monolayers of thiolates on metals as a form of nanotechnology. Chemical Reviews 105: 1103-1169.
2. Hohlfeld J, Wellershoff SS, Gudde J, Conrad U, Jahnke V, et al. (2000) Electron and lattice dynamics following optical excitation of metals. Chemical Physics 251: 237-258.
3. Frischkorn C, Wolf M (2006) Femtochemistry at metal surfaces: nonadiabatic reaction dynamics. Chemical Reviews (Washington, D C) 106: 4207-4233.
4. Wang Z, Carter JA, Lagutchev A, Koh YK, Seong NH, et al. (2007) Ultrafast flash thermal conductance of molecular chains. Science 317: 787-790.
5. Eisenthal KB (1996) Liquid interfaces probed by second-harmonic and sum-frequency spectroscopy. Chemical Reviews 96: 1343-1360.
6. Wang ZH, Cahill DG, Carter JA, Koh YK, Lagutchev A, et al. (2008) Ultrafast dynamics of heat flow across molecules. Chemical Physics 350: 31-44.
7. Zhang Y, Barnes GL, Yan T, Hase WL (2010) Model non-equilibrium molecular dynamics simulations of heat transfer from a hot gold surface to an alkylthiolate self-assembled monolayer. Physical Chemistry Chemical Physics 12: 4435-4445.
8. Manikandan P, Carter JA, Dlott DD, Hase WL (2011) Effect of Carbon Chain Length on the Dynamics of Heat Transfer at a Gold/Hydrocarbon Interface: Comparison of Simulation with Experiment. Journal of Physical Chemistry C 115: 9622-9628.
9. Lagutchev AS, Patterson JE, Huang WT, Dlott DD (2005) Ultrafast dynamics of self-assembled monolayers under shock compression: Effects of molecular and substrate structure. Journal of Physical Chemistry B 109: 5033-5044.
10. Hirose C, Akamatsu N, Domen K (1992) Formulas for the Analysis of the Surface Sfg Spectrum and Transformation Coefficients of Cartesian Sfg Tensor Components. Applied Spectroscopy 46: 1051-1072.
11. Lambert AG, Davies PB, Neivandt DJ (2005) Implementing the theory of sum frequency generation vibrational spectroscopy: A tutorial review. Applied Spectroscopy Reviews 40: 103-145.
12. Cho M (2009) Two-dimensional optical spectroscopy. Boca Raton: CRC Press.
13. Hamm P, Zanni MT (2011) Concepts and Methods of 2D Infrared Spectroscopy. New York: Cambridge University Press.
14. Carter JA, Wang ZH, Dlott DD (2009) Ultrafast Nonlinear Coherent Vibrational Sum-Frequency Spectroscopy Methods To Study Thermal Conductance of Molecules at Interfaces. Accounts of Chemical Research 42: 1343-1351.
15. Bordenyuk AN, Benderskii AV (2005) Spectrally- and time-resolved vibrational surface spectroscopy: ultrafast hydrogen-bonding dynamics at D2O/CaF2 interface. Journal of Chemical Physics 122: 134713.
16. Herzberg G (1945) Infrared and Raman Spectra of Polyatomic Molecules. New York: Van Nostrand. 197 p.
17. Guo Z, Zheng W, Hamoudi H, Dablemont C, Esaulov VA, et al. (2008) On the chain length dependence of CH3 vibrational mode relative intensities in sum frequency generation spectra of self assembled alkanethiols. Surface Science 602: 3551-3559.
18. Carter JA, Wang ZH, Fujiwara H, Dlott DD (2009) Ultrafast Excitation of Molecular Adsorbates on Flash-Heated Gold Surfaces. Journal of Physical Chemistry A 113: 12105-12114.
19. Lagutchev A, Hambir SA, Dlott DD (2007) Nonresonant background suppression in broadband vibrational sum-frequency generation spectroscopy. Journal of Physical Chemistry C 111: 13645-13647.

20. Suarez C, Bron WE, Juhasz T (1995) Dynamics and transport of electronic carriers in thin gold films. Physical Review Letters 75: 4536-4539.
21. Nihonyanagi S, Eftekhari-Bafrooei A, Borguet E (2011) Ultrafast vibrational dynamics and spectroscopy of a siloxane self-assembled monolayer. Journal of Chemical Physics 134: 084701.
22. Laaser JE, Xiong W, Zanni MT (2011) Time-Domain SFG Spectroscopy Using Mid-IR Pulse Shaping: Practical and Intrinsic Advantages (vol 115, pg 2536, 2011). Journal of Physical Chemistry B 115: 9920-9920.
23. Fujiwara H, Brown KE, Dlott DD (2010) High-energy flat-top beams for laser launching using a Gaussian mirror. Applied Optics 49: 3723-3731.
24. Patterson JE, Dlott DD (2005) Ultrafast shock compression of self-assembled monolayers: A molecular picture. Journal of Physical Chemistry B 109: 5045-5054.
25. Born M, Wolf E, Bhatia AB (1999) Principles of Optics: Electromagnetic Theory of Propagation, Interference and Diffraction of Light. Cambridge University Press. pp. 735-789.
26. Lynch DW, Hunter WR (1997) Comments on the Optical Constants of Metals and an Introduction to the Data for Several Metals. In: Edward DP, editor. Handbook of Optical Constants of Solids. Burlington: Academic Press. pp. 275-367.
27. Wilson RB, Apgar BA, Martin LW, Cahill DG (2012) Thermoreflectance of metal transducers for optical pump-probe studies of thermal properties. Optics Express 20: 28829-28838.
28. Kimura Y, Yamamoto Y, Fujiwara H, Terazima M (2005) Vibrational energy relaxation of azulene studied by the transient grating method. I. Supercritical fluids. Journal of Chemical Physics 123: 054512.
29. Fujiwara H, Terazima M, Kimura Y (2008) Transient grating study on vibrational energy relaxation of bridged azulene–anthracene's. Chemical Physics Letters 454: 218-222.
30. Fateley WG, Miller FA (1963) Torsional Frequencies in the Far Infrared .3. The Form of the Potential Curve for Hindered Internal Rotation of a Methyl Group. Spectrochimica Acta 19: 611-628.
31. Sung J, Kim D (2007) Motional effect on the sum-frequency vibrational spectra from methanol surface. Journal of the Korean Physical Society 51: 145-148.
32. Wei X, Shen YR (2001) Motional effect in surface sum-frequency vibrational spectroscopy. Physical Review Letters 86: 4799-4802.
33. Daragan VA, Mayo KH (1997) Motional model analyses of protein and peptide dynamics using C-13 and N-15 NMR relaxation. Progress in Nuclear Magnetic Resonance Spectroscopy 31: 63-105.
34. Wittebort RJ, Szabo A (1978) Theory of Nmr Relaxation in Macromolecules - Restricted Diffusion and Jump Models for Multiple Internal Rotations in Amino-Acid Side-Chains. Journal of Chemical Physics 69: 1722-1736.
35. Hiyama Y, Roy S, Cohen JS, Torchia DA (1989) Solid-State H-2 Nmr-Study of Thymidine - Base Rigidity and Ribose Ring Flexibility in Deoxynucleosides. Journal of the American Chemical Society 111: 8609-8613.
36. Somorjai GA, Frei H, Park JY (2009) Advancing the frontiers in nanocatalysis, biointerfaces, and renewable energy conversion by innovations of surface techniques. Journal of the American Ceramic Society 131: 16589-16605.
37. Yamaguchi S, Tahara T (2008) Heterodyne-detected electronic sum frequency generation: "up" versus "down" alignment of interfacial molecules. journal of chemical physics 129: 101102.

TABLE

Table 1: Kinetic parameters [‡]

| $I_0$ [%] | $A_1$ [%] | $\tau_1$ [ps] | $A_2$ [%] | $\tau_2$ [ps] |
|---|---|---|---|---|
| 75±7 | 10±1 | 5.2±0.7 | 15±4 | 5.2±0.7 |
| 92.0±0.9 | 8.0±0.7 | 5.7±1.2 | - | - |

[‡] Error are standard deviations.

FIGURES

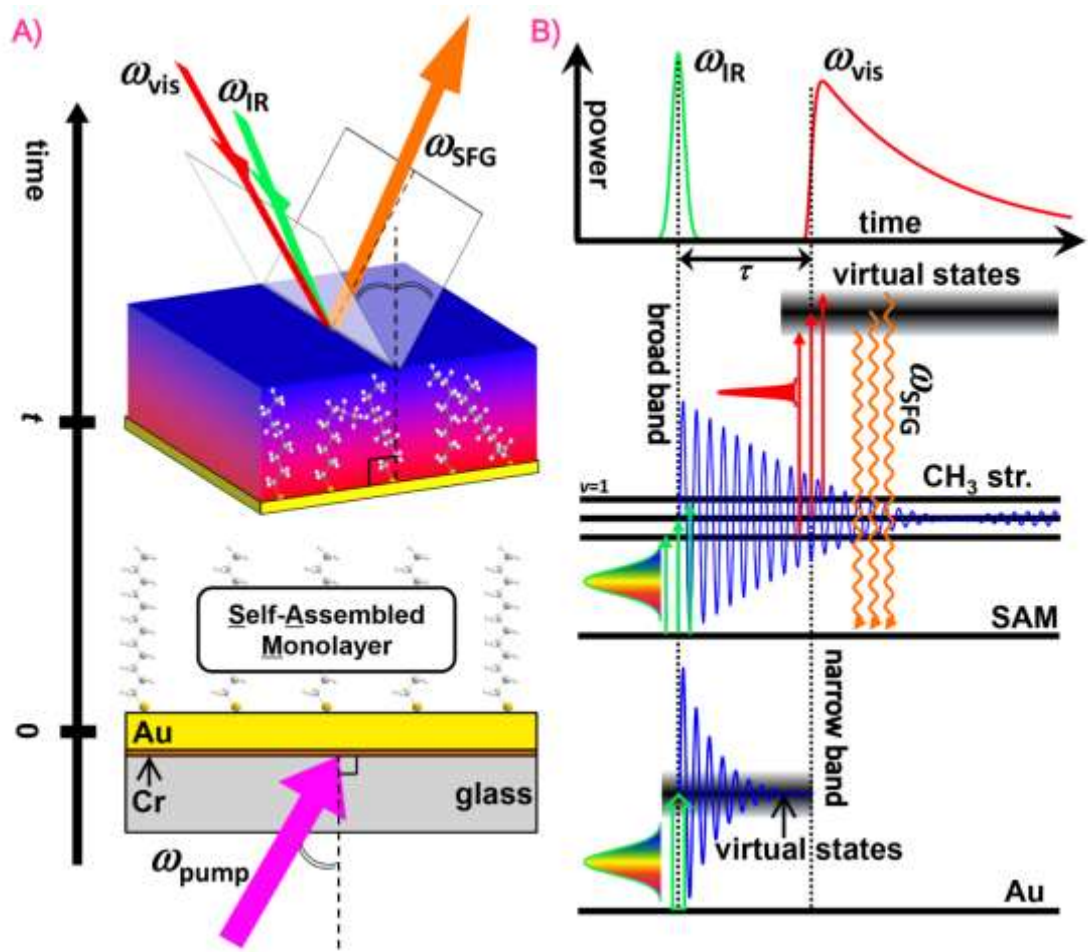

**Figure 1.** (a) The geometry of time-resolved vibrational sum-frequency generation. The pump pulse ($\omega_{pump}$) heated thin metal layers from the back side. Two probe pulses ($\omega_{IR}$ and $\omega_{vis}$) were used to measure the vibrational spectrum of the self-assembled monolayer via sum frequency generation ($\omega_{SFG}$) after a time delay $t$ of $\omega_{IR}$ pulse from $\omega_{pump}$ pulse. (b) The principle of our background-suppression technique. The two probe pulses arrived with a time delay $\tau$. The first pulse $\omega_{IR}$ induced the first order polarization on both the SAM and Au surface. The second pulse $\omega_{vis}$ interacted only with SAM polarization to induce a stimulated Raman process $\omega_{SFG}$.

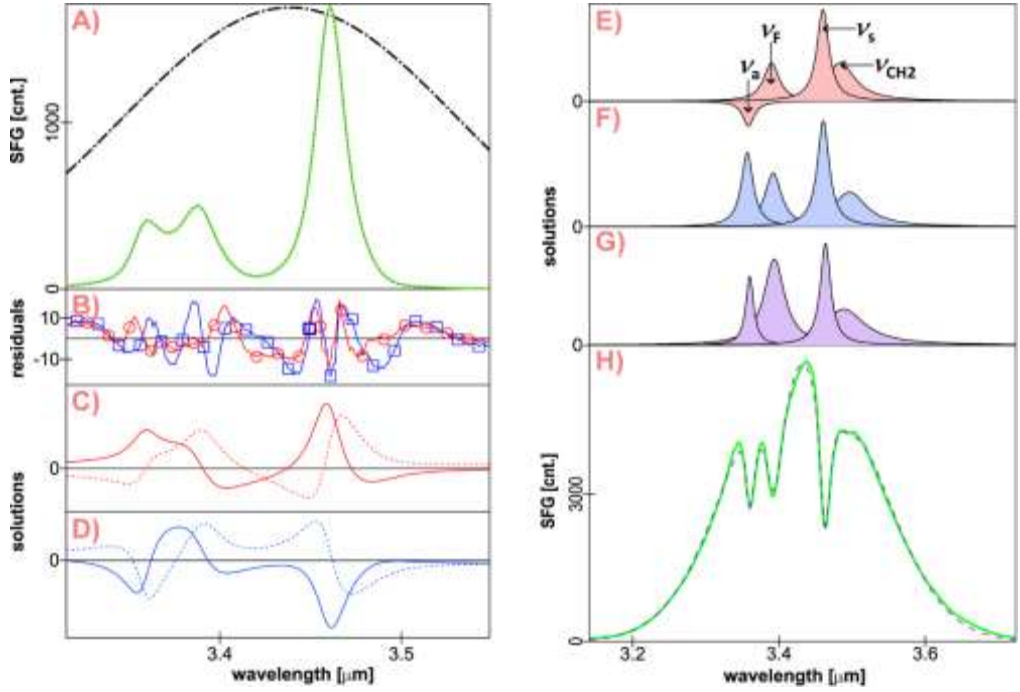

**Figure 2.** (a) The background-suppressed vibrational sum-frequency generation (SFG) spectrum in the C-H stretching region (green solid curve in the first graph) without the pump pulse $\omega_{pump}$. The horizontal axis shows the corresponding vibrational frequency (201 pixels). The spectral function (eq. 1) for the first solution (red dash) and Gaussian $g(\omega_{IR})$ optimized to the standard SFG spectrum in (h) (black double dash) are also shown. (b) Residuals of the first and second solutions (red circles and blue squares, respectively). (c)/(d) The real (solid) and imaginary (dashed) parts of $\chi_2$ resonant amplitudes that appeared in the first (c, $\tau = 0.9$ ps) and second (d, $\tau = 0.6$ ps) solutions, respectively. Fermi's resonance ($\nu_F$) term was defined to have a zero phase ($\phi_F = 0$), since the absolute phase values were undetermined. (e)/(f)/(g) Lorentzian function corresponding to each resonant $\chi_2$ term with the sign of amplitude $A_n$ obtained from fitting to the first ((e)) and second ((f)) phase-constrained solutions for the back-ground suppressed spectrum as well as the standard spectrum ((g)), respectively. These amplitudes have arbitrary absolute signs. (h) A standard SFG spectrum without background suppression (green solid) and eq. 1 (purple dashed).

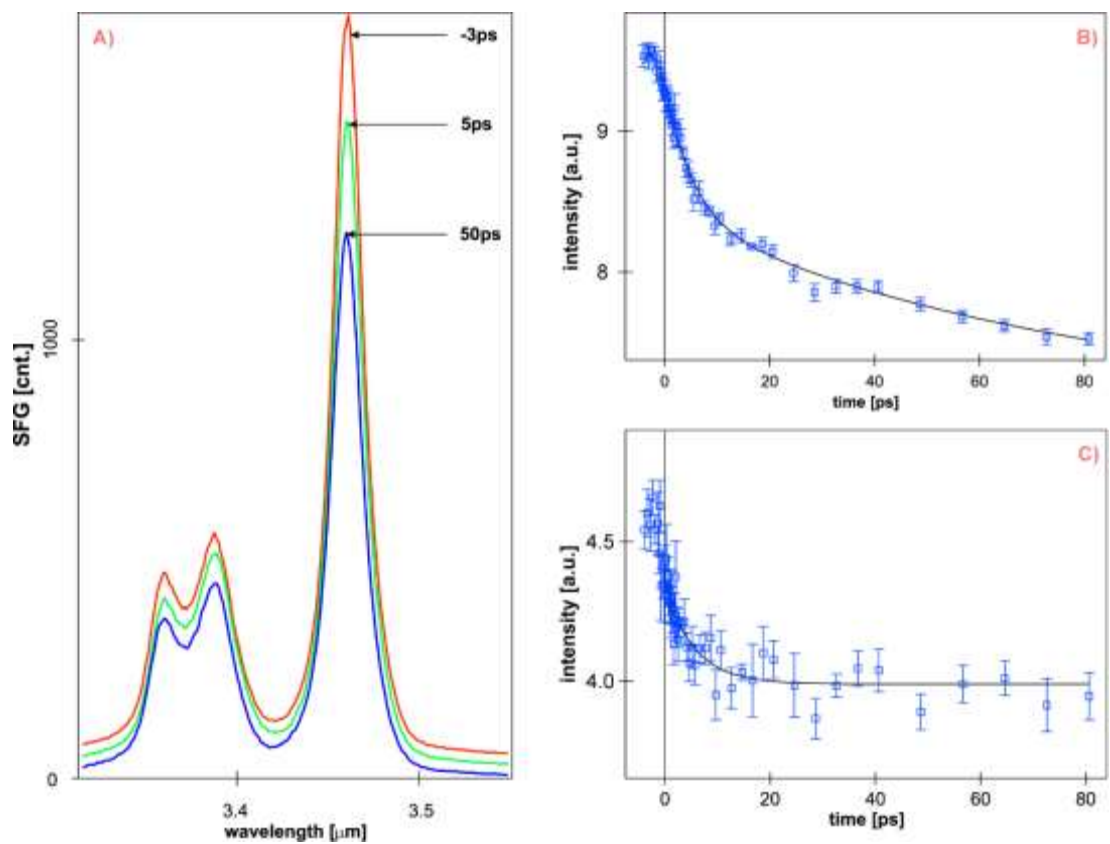

**Figure 3.** (a) Averaged time-resolved background-suppressed vibrational SFG spectra in the C-H stretching region. The selected delay times are -3, 5, and 50 ps from the upper to lower curves. (b)/(c) Methyl C-H stretching intensity time profiles for the second (blue square) solutions (eq. 1) with standard errors. The symmetric ($\nu_s$) and asymmetric ($\nu_a$) modes are shown in (b) and (c), respectively. Each data point is the average of seven global-fit results from each scan after eliminating spikes. The optimized double and single exponential curves for $\nu_s$ and $\nu_a$ modes of the second solution are also shown (black curves), respectively.

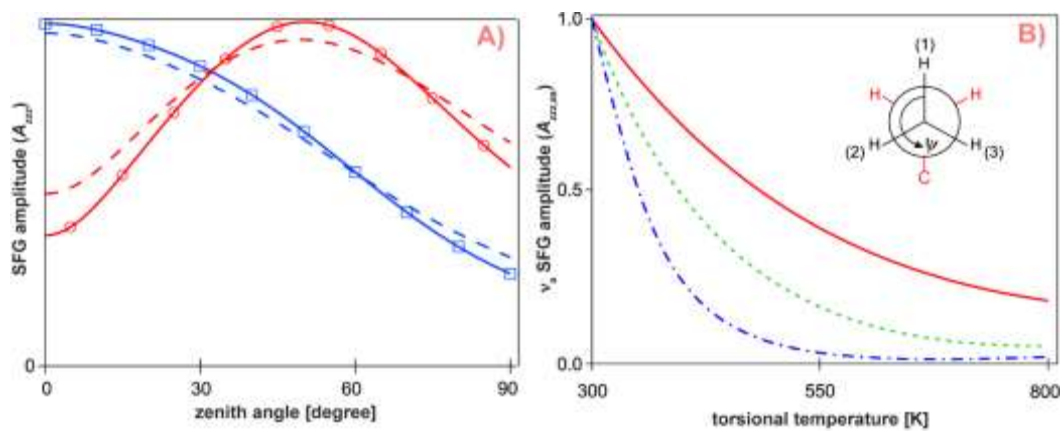

**Figure 4.** (a) Resonant SFG amplitudes of methyl C-H symmetric ($A_{zzz,s}$, red circle) and asymmetric ($A_{zzz,as}$, blue square) vibrations at the slow limit as the function of zenith angle $\theta$, provided a normal distribution with a 16° (solid) and 20° (dashed) standard deviations. The fast limit of $\nu_s$ amplitude agrees with the slow limit within the size of makers (blue squares), while $\nu_a$ signal vanishes due to uniform distribution of torsion angle $\psi$. The inset shows the definitions of $\theta$, as well as $\psi$. (b) Temperature effects of the $\nu_a$ SFG amplitude $A_{zzz,as}$ normalized to 300 K with $\tau = 200$ (red solid), 400 (green dash), and 800 fs (blue double-dash) delay time (see text). The inset illustrates the three local minima (1-3) designated as conformers in the Site Jump model.